%% file: main.tex
\documentclass[journal]{IEEEtran}

\usepackage{graphicx}
\usepackage{url}
\usepackage{amsmath}
\usepackage{amssymb}
\usepackage{cases}
\usepackage[ruled,vlined]{algorithm2e}
\usepackage{subcaption}
\usepackage{color}

\newtheorem{proposition}{{\bf Proposition}}

\def\done{\hspace*{\fill} \rule{1.8mm}{2.5mm} \\ }

\begin{document}
\title{Modeling and Quantifying the Forces Driving Online Video Popularity Evolution}

\author{
\IEEEauthorblockN{Jiqiang Wu, Yipeng Zhou, Dah Ming Chiu\\}
\IEEEauthorblockA{
Email: jqwu@tcl.com; yipeng.job@gmail.com; dmchiu@ie.cuhk.edu.hk}
}

\maketitle

\begin{abstract}
Video popularity is an essential reference for optimizing resource allocation and video recommendation in online video services.
However, there is still no convincing model that can accurately depict a video's popularity evolution. In this paper, we propose a dynamic popularity model by modeling the video information diffusion process driven by various forms of recommendation. Through fitting the model with real traces collected from a practical system, we can quantify the strengths of the recommendation forces. Such quantification can lead to characterizing video popularity patterns, user behaviors and recommendation strategies, which is illustrated by a case study of TV episodes.
\end{abstract}

\input ./sec/introduction.tex
\input ./sec/assumption.tex

\input ./sec/model.tex

\input ./sec/evaluate.tex
\input ./sec/cstv.tex
\input ./sec/related.tex

\input ./sec/conclusion.tex

\bibliographystyle{IEEEtran}
\bibliography{sigproc}

\input ./sec/appendix.tex

\end{document}

%% file: sec/introduction.tex
\section{Introduction}
\label{Sec:Introduction}

Video popularity is an essential reference for optimizing resource allocation and video recommendation in online video services. Over the past few years researchers have studied several aspects of video popularity such as popularity distributions~\cite{chaTON09}, evolution patterns~\cite{yipengTMM15}, prediction~\cite{pintoWSDM13}\cite{ahmedWSDM13}, models~\cite{avramovaICEI09}\cite{jasonIWQOS15} and so on. However, there is still no convincing model that can accurately depict a video's popularity evolution. Such a dynamic model is useful not only for predicting popularity evolution, but also for characterizing user behaviors and content providers' recommendation strategies.

Large-scale online video providers provide millions of videos such that users heavily rely on recommendations to watch their favored videos (besides ad-hoc searching). As a result, a video's popularity evolution is largely affected by how it is recommended to users. Direct and word-of-mouth (WOM) recommendations are the two major mechanisms which are repeatedly observed and discussed by previous works~\cite{szaboCACM10}\cite{liuNOSSDAV12}\cite{jasonIWQOS15}. Direct recommendation means that videos are exposed to users typically through websites' front pages, TV promotion channels, advertisements, etc and WOM means that videos are shared through various social networks such as Facebook, Twitter, BBS forums, emails and even offline social networks like colleges and relatives.

The two recommendation mechanisms result in an information diffusion process. By modeling this process we can derive the incremental user population of watching a video and thus derive the video's popularity evolution. However, the complication of such a model lies in the fact that practical recommendation resources are limited. For example, because of limited positions in front-page promotion, existing videos can only be recommended for a certain period as new videos are constantly brought in. So far there is little knowledge of how the two recommendations drive popularity dynamics with limited resources, which will be discussed by this study.

On the other hand, given a video's observed popularity evolution, we can fit it to the model to reveal and quantify its spreading process and the driving forces of recommendation. Such quantification will provide a systematic approach to characterizing a dynamic video system by a set of insightful parameters.
Both academia and industry can benefit from this approach as it can be used to detect latent user behaviors, identify inadequately recommended videos and evaluate recommendation strategies. We will illustrate it through a case study of TV episodes in Section~\ref{Sec:cstv}.

%% file: sec/assumption.tex
\section{Definitions and Assumptions}
\label{Sec:Assumption}

Video popularity is measured by view count throughout this paper. Without loss of generality, we discuss the popularity evolution for a particular video.
Let $x(t)$ be the \emph{cumulative} popularity evolution function, i.e the number of users who have watched the video up to time $t$. Then the (instant) popularity at time $t$ is $x'(t)$.  In our model, all recommendations are classified as either direct recommendation or WOM recommendation. The recommendation made by content creators, providers and any other entity for their own benefits, is regarded as direct recommendation; while the recommendation made by users based on their own interests is regarded as WOM recommendation. We define the intrinsic attractiveness $q$ of a video as the probability that the video is favored by any user. As users learn the video's existence through recommendations, only an average fraction $q$ of them will watch the video.

To simplify the analysis, we make the following assumptions. Firstly, users never replay a video. This assumption is supported by some previous work~\cite{yipengTMM15}\cite{liuNOSSDAV12}. Secondly, users make independent decision to select videos. It may not be true for related videos. We will introduce a technique in Sec.~\ref{Sec:cstv} to tackle this issue. Thirdly, for a given video, the total potential user population that can be recommended is fixed. In practice, this assumption holds for a relatively short period, e.g. several months. Finally, users watch the video immediately after they learn its existence and are interested in it. Some previous work assume there is a user reaction process. However, for popular videos this process can be ignored, as an approximation~\cite{jasonIWQOS15}.

%% file: sec/model.tex
\section{Model}\label{Sec:model}

The information diffusion process may be driven by both direct recommendation and WOM recommendation, but it is complicated to involve both of them in a single process. A previous work~\cite{jasonIWQOS15} has shown that for a given video either direct or WOM recommendation plays the main driving role. Therefore we study two simplified information diffusion processes: \emph{DModel} driven by direct recommendation, and \emph{WModel} driven by WOM recommendation.

Without loss of generality, we model a particular video's popularity evolution by assuming users make independent video selection as stated in Sec.~\ref{Sec:Assumption}.
We create a fluid epidemic model to depict its information diffusion process.
Let $t$ denote the elapsed time since the video is born (made available online). At time $t$, $\mathcal{S}(t)$ is the set of users who do not know the existence of the video. $\mathcal{X}(t)$ is the set of users who know the existence of and are interested in the video, and $\mathcal{Y}(t)$ is the set of users who know the existence of but are \emph{not} interested in the video. Let $s(t)$, $x(t)$ and $y(t)$ be the cardinality of the sets $\mathcal{S}(t)$, $\mathcal{X}(t)$ and $\mathcal{Y}(t)$, respectively. Under the assumption of fixed population $N$, we have $s(t)+x(t)+y(t)=N$, and approximately $\frac{x(t)}{y(t)}=\frac{q}{1-q}$.  Finally, let $v(t)=x'(t)$ be the dynamic view count, i.e. the (instant) popularity at time $t$.

\subsection{DModel}
Because of limited recommendation resource in practice, it is impossible that the video information can be diffused to all users instantly. Therefore we define $\alpha$ as the direct recommendation rate, i.e. the rate that users flow from set $\mathcal{S}(t)$ to $\mathcal{X}(t) \cup \mathcal{Y}(t)$. Note that the total potential user population $N$ is typically a very large number, e.g. several tens of millions. Thus we can create a fluid model as follows:
\begin{eqnarray}
  \label{EQ:dx_alpha}
  x'(t) & = &\alpha q s(t),\\
  y'(t) & = &\alpha (1-q) s(t),
\end{eqnarray}
with initial condition $x(0)=y(0)=0$ and $s(0)=N$. Together with the equation $s(t)+x(t)+y(t)=N$, we can derive that
\begin{eqnarray}
  \label{EQ:xt_alpha}
  x(t) &=& qN(1-e^{-\alpha t}) \\
  \label{EQ:dxt_alpha}
  x'(t) &=& \alpha qNe^{-\alpha t}
\end{eqnarray}
There is no restriction on direct recommendation period in the above process. In practice, as new videos are constantly brought into the system, each existing video is only promoted for a certain period, depending on the estimated video popularity, user feedback, arrival rate of new videos and so on.
Thus we introduce another parameter $t_e$ as the time when the direct recommendation aborts. After time $t_e$, no new user will join in $\mathcal{X}(t)$ and then $x'(t)$ will decrease to $0$ very fast with rate $\gamma \gg \alpha$.

Summarizing the above discussion, we derive the video popularity evolution driven by direct recommendation as
\begin{eqnarray}
v(t)=\left\{
\begin{aligned}
 & \alpha qNe^{-\alpha t}, & 0 < t \leq t_e \\
 & \alpha qNe^{-\alpha t_e} \cdot e^{-\gamma(t-t_e)}, & t > t_e
\end{aligned}\nonumber
\right.
\end{eqnarray}

\subsection{WModel}
In the WModel we split the timeline into time slots. Let $\mathcal{D}(t)$ denote the set of \emph{new} users who join the set $\mathcal{X}(t)$ in time slot $t$, and let $\Delta x(t) = x(t)-x(t-1)$ be the cardinality of $\mathcal{D}(t)$. In time slot $t$, the users in $\mathcal{D}(t)$ perform two operations: a) finish watching the video and thus generate $\Delta x(t)$ view count, and b) recommend the video to other users \emph{merely} in this time slot. In other words, users recommend a video immediately after they finish watching it and the recommendation period only lasts one time slot. This is consistent with the reality. User attentions are often attracted by new videos such that no users will keep recommending a video for a long period. The granularity of time slot can be one day or one week depending on the scenario.

Let $\beta$ be the average number of friends a new user will recommend the video, normalized by $N$. Then the WOM recommendation process is a discrete time model:

\begin{eqnarray}
  \label{EQ:dx_beta}
  \Delta x(t+1)& =& \beta q \Delta x(t)s(t),\\
  y(t+1) & = & \frac{x(t+1)(1-q)}{q}.
\end{eqnarray}
The interpretation of the above equations is as follows. The new users joining the set $\mathcal{X}(t+1)$ in time slot $t+1$, i.e. the $\Delta x(t+1)$ users in $\mathcal{D}(t+1)$, are attracted by WOM recommendation made by the $\Delta x(t)$ users in $\mathcal{D}(t)$.  $\beta N$ is the average number of users who will be recommended by a user and $\frac{s(t)}{N}$ is the probability that the video information is unknown for any user\footnote{This means a user recommends the video to random $\beta N$ users among the population without keeping track of who have watched the video.}.  Thus, $\beta N\times \frac{s(t)}{N} \times q$ (only a fraction $q$ are interested) is the number of new users contributed by a user in $\mathcal{D}(t)$. Given $\Delta x(t)$ users recommending the video, we have $\Delta x(t+1) = \beta q \Delta x(t)s(t)$.

It is hard to get the closed-form solution of $x(t)$ or $v(t)$. To obtain its theoretical insights, we consider an approximate continuous model by letting $x''(t) \approx \frac{\Delta x(t+1) - \Delta x(t)}{\Delta t}$ and $x'(t) \approx \frac{\Delta x(t)}{\Delta t}$. Then Eq.~\ref{EQ:dx_beta} becomes
\begin{equation}\label{EQ:dx_beta_cont}
  x''(t) = x'(t)[\beta q s(t) - 1]
\end{equation}

In order for the WOM to proceed, we should have an initial condition $x(0) = x_0 > 0$. $x_0$ is the number of seeds and we allow $x_0 < 1$. The period with $x(t) < 1$ indicates how long it takes for the first user to discover the video.

Given the fixed total user population $s(t)+\frac{x(t)}{q}=N$ (recall $\frac{x(t)}{y(t)}=\frac{q}{1-q}$) and initial conditions that $x(0)=x_0$, $x'(0) \approx x(1)-x(0) = \beta q x_0 (N-\frac{x_0}{q})$, we can get
\begin{equation}\label{EQ:xt_beta}
  x(t) = \frac{x_1 + g(t)x_2}{1+g(t)}
\end{equation}

Here $x_1 = qN+\frac{\varphi-1}{\beta}$, $x_2 = qN+\frac{-\varphi-1}{\beta}$,
$g(t) = \frac{x_1-x_0}{x_0-x_2}e^{-\varphi t}$ and $\varphi = \sqrt{(\beta qN-1)^2-(\beta x_0-1)^2+1}$. The detailed derivation is in \cite{appendix}.

The solution of $x(t)$ is complicated for deriving the evolution of $v(t)$. Instead, we focus on analyzing the limit of $x(t)$. Clearly, as $t$ approaches infinity, $g(t)$ approaches $0$ and $x(t)$ approaches $x_1$. In other words, the limit of $x(t)$ depends on the value of $x_1$, which should never exceed $qN$. The gap $qN - x_1$ is essential as it indicates whether the video has been sufficiently recommended.

\begin{proposition}
\label{Prop:PropRecCond}
If $\beta\geq \frac{2}{qN+x_0}$, the video information can be diffused to all users; otherwise, only a subset of users know the video information.
\end{proposition}
The video information is diffused to all users if and only if $x_1\geq qN$. It is not difficult to derive the condition $\beta<\frac{2}{qN+x_0}$ by letting $x_1<qN$. The detailed proof is omitted. From Proposition~\ref{Prop:PropRecCond}, we can see that either small $\beta$ or $x_0$ can result in insufficient recommendation\footnote{Note, if $\beta \geq\frac{2}{qN+x_0}$, it is possible that $x_1>qN$, incurred by the approximation error of the continuous model. This error is absent from the discrete model, which is used for evaluation and curve fitting.}.

\begin{figure}[ht]
\centering
\includegraphics[width=0.48\textwidth]{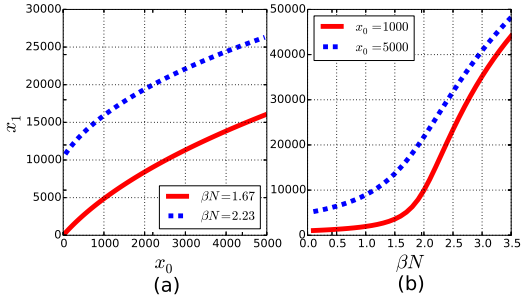}
\caption{ Final user population $x_1$ increases as a concave function of $x_0$. Here $N =100,000$ and $q=0.5$. }
\label{Fig:x1_x0_beta}
\end{figure}

In the following discussion, we focus on the case where $x_1<qN$. We are interested in how $x_0$ and $\beta$ affect $x_1$, i.e. the number of users who finally watch the video.
\begin{proposition}
\label{Prop:x1_x0}
If $\beta< \frac{2}{qN+x_0}$, the final user population $x_1$ increases as a concave function of $x_0$.
\end{proposition}
This proposition can be proved from $\frac{\partial x_1}{\partial x_0} = \frac{-(\beta x_0 - 1)}{\beta} > 0$ and $\frac{\partial^2 x_1}{\partial x_0^2}= \frac{-\beta[(\beta qN-1)^2+1]}{\varphi^3} < 0$.
Fig.~\ref{Fig:x1_x0_beta}a shows how $x_0$ affects final population $x_1$ under different values of $\beta$.
We can see that $x_1$ only increases from $5000$ to $15000$ as $x_0$ increases from $1000$ to $5000$ when $\beta N=1.67$, indicating the inefficiency for diffusing information by increasing $x_0$
\begin{proposition}
\label{Prop:x1_beta}
There exists a $\frac{1}{qN+x_0} < \theta < \frac{2}{qN+x_0}$, such that
if $\beta \leq \theta$ the final user population $x_1$ increases as a convex function of $\beta$;
while if $\theta \leq \beta < \frac{2}{qN+x_0}$ the final user population $x_1$ increases as a concave function of $\beta$ .
\end{proposition}
The proof is in ~\cite{appendix}. Fig.~\ref{Fig:x1_x0_beta}b shows how $x_1$ increases as $\beta$ increases under different values of $x_0$.
There exists a threshold such that if $\beta$ is larger than the threshold, it is very effective to lift the final user population $x_1$.
However, in real world, it is much more difficult for video providers to control $\beta $, because $\beta $ is mainly determined by individual  users.

\subsection{Discussion}

Although we have derived the DModel and WModel based on different driving forces of recommendation, for a given video it is unknown which model fits it best. If the view count trace of the video can be obtained, the type of information diffusion process is decided by the model with smaller fitting error. In addition, through fitting the video traces with the theoretical models, we can quantify the strength of each recommendation force, which is helpful for evaluating recommendation strategies and observing the user behavior. The details will be presented in Sec.~\ref{Sec:cstv}.

The assumption that a video's information is diffused either by direct recommendation or WOM recommendation may not be valid, because both forces can simultaneously drive the process.
However, it is reported in~\cite{jasonIWQOS15} that for most videos only one force plays the major role. Therefore, we can merely consider one single force to simplify the information diffusion process. With the simplification, the limitation of recommendation resources (e.g. $t_e$ and short WOM period) can be studied, which is a significant advantage of our model over the model proposed in~\cite{jasonIWQOS15} that analyzed multiple forces simultaneously.

%% file: sec/evaluate.tex
\section{Evaluation}
In this section, we verify the DModel and WModel by fitting them with the view count traces collected from a practical system. Each video will be fitted with both models, but only the one with smaller error is used for evaluation. For convenience, we refer to the model with better fitting as \emph{BModel}.

The \emph{normalized mean square error}(NMSE) is used as the metric to evaluate fitting error, which is defined as
\begin{eqnarray}
  \label{EQ:NMSE}
  NMSE &=& \frac{\frac{1}{T}\sum_{i=1}^{T}(\hat{v}_t-v_t)^2}{\big(\frac{\sum_{i=1}^{T}v_t}{T}\big)^2},
\end{eqnarray}
where $\hat{v}_t$ and $v_t$ represent the $t^{th}$ day's view count calculated by the model and collected from the real system, respectively. Note that NMSE has been normalized by the square of the real trace's average view count so that we can compare cases with different total view count.

\subsection{Dataset}
Instead of evaluating all videos, we focus on the videos from the four most important types: Movie, TV, News and Music Video (MV). We collected all videos of the four types that are uploaded between September 1, 2014 to January 31, 2015, from the system's viewing records. Each viewing record contains the following information: time, user id and video id. For each video, we collected its daily view count trace for six months. The videos with less than $1000$ total view count are removed from evaluation because their traces are likely to be driven by occasional views hence are too noisy. Removing cold videos does not affect the generality of this study since their view count takes less than $3\%$ of the total view count. The resulting dataset contains 1469 movies, 9705 TV episodes, 30720 News and 4736 MVs.

Before fitting the models with the trace dataset, we need to figure out the total user population for each video type. We cannot simply take the number of total user id as the population because it would include occasional viewers. Instead, we consider the relatively active users who have watched a fair number of videos. Thus, for each video type we rank all users according to their view count in increasing order. Then the users with few view count are excluded until 25\% total view count are removed, along with their contributed views to the videos\footnote{The aforementioned procedure of removing cold videos are done after fixing total user population for each video type}. The number of remaining users varies from around 44 millions to around 81 millions\footnote{Due privacy issue, we don't report the actual user number.}. Although this is a heuristic rule to exclude occasional users, one does not have to exactly get the value $N$, which is mainly used to normalize the other parameters $\alpha$, $\beta$, etc. As long as the same $N$ is used for a video type, the comparison is fair for videos within the same type.

\subsection{Model fitting}

\begin{figure}[ht]
\centering
\includegraphics[width=0.48\textwidth]{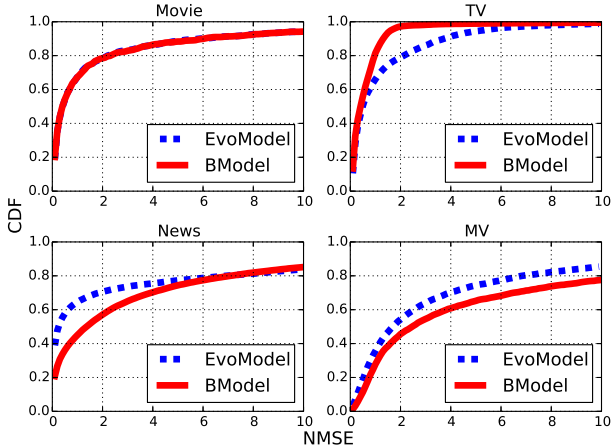}
\caption{ Fitting error distributions of the four video types. }
\label{Fig:nmse}
\end{figure}

The Levenberg-Marquardt algorithm, a common algorithm for finding least square error, is used to search the optimal parameters when fitting the models with trace data. The NMSE for each video is calculated by Eq.~\ref{EQ:NMSE} with the optimal parameters. We use the EvoModel proposed in~\cite{jasonIWQOS15} as the benchmark and compare the cumulative distribution functions (CDF) of NMSE. The results are plotted in Fig.~\ref{Fig:nmse}. Note that the NMSE of BModel is the smaller NMSE of DModel and WModel. Through model fitting, we automatically classify the videos into two groups: d-recommended videos and w-recommended videos.

From Fig.~\ref{Fig:nmse} we can see that for Movie and TV, BModel achieves better performance than EvoModel (the two CDF curves of Movie almost overlap), while EvoModel slightly outperforms BModel for News and MV. We explain these results as follows. EvoModel analyzes multiple forces simultaneously without restricting each recommendation resource ($t_e$ and short WOM period). According to ~\cite{jasonIWQOS15}, the information of most movies and TV episodes is diffused either by direct recommendation or WOM recommendation. Thus BModel can achieve better fitting results by incorporating the limitation of recommendation resources. However, for News and MV, there exists a number of videos relying on both recommendation forces, which results in worse fitting results for BModel. In fact, only those warm videos are prone to be affected by both forces. BModel is better for most popular videos.

%


%% file: sec/cstv.tex
\section{Case Study: TV}\label{Sec:cstv}
As we have discussed, our dynamic popularity model can quantify the strengths of the recommendation forces through model fitting so that we can observe user behaviors and evaluate recommendation strategies. We conduct case study with TV episodes to illustrate this point.

\subsection{Aggregating episodes}
\begin{figure*}[ht]
\centering
        \begin{subfigure}[b]{0.30\linewidth}
                \includegraphics[width=1.0\linewidth]{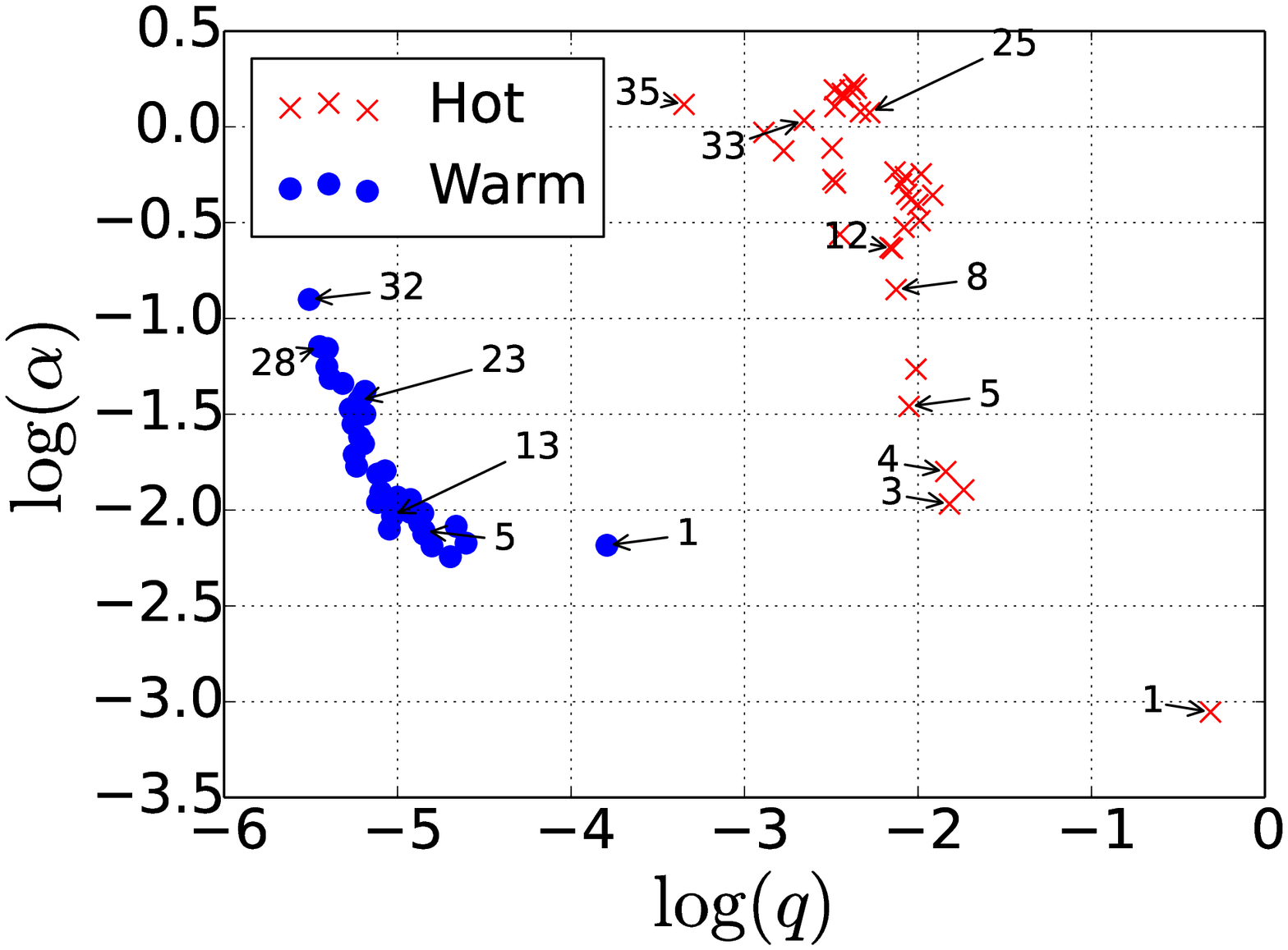}
                \caption{}
                \label{fig:cstv1}
        \end{subfigure}
        \begin{subfigure}[b]{0.30\linewidth}
                \includegraphics[width=1.0\linewidth]{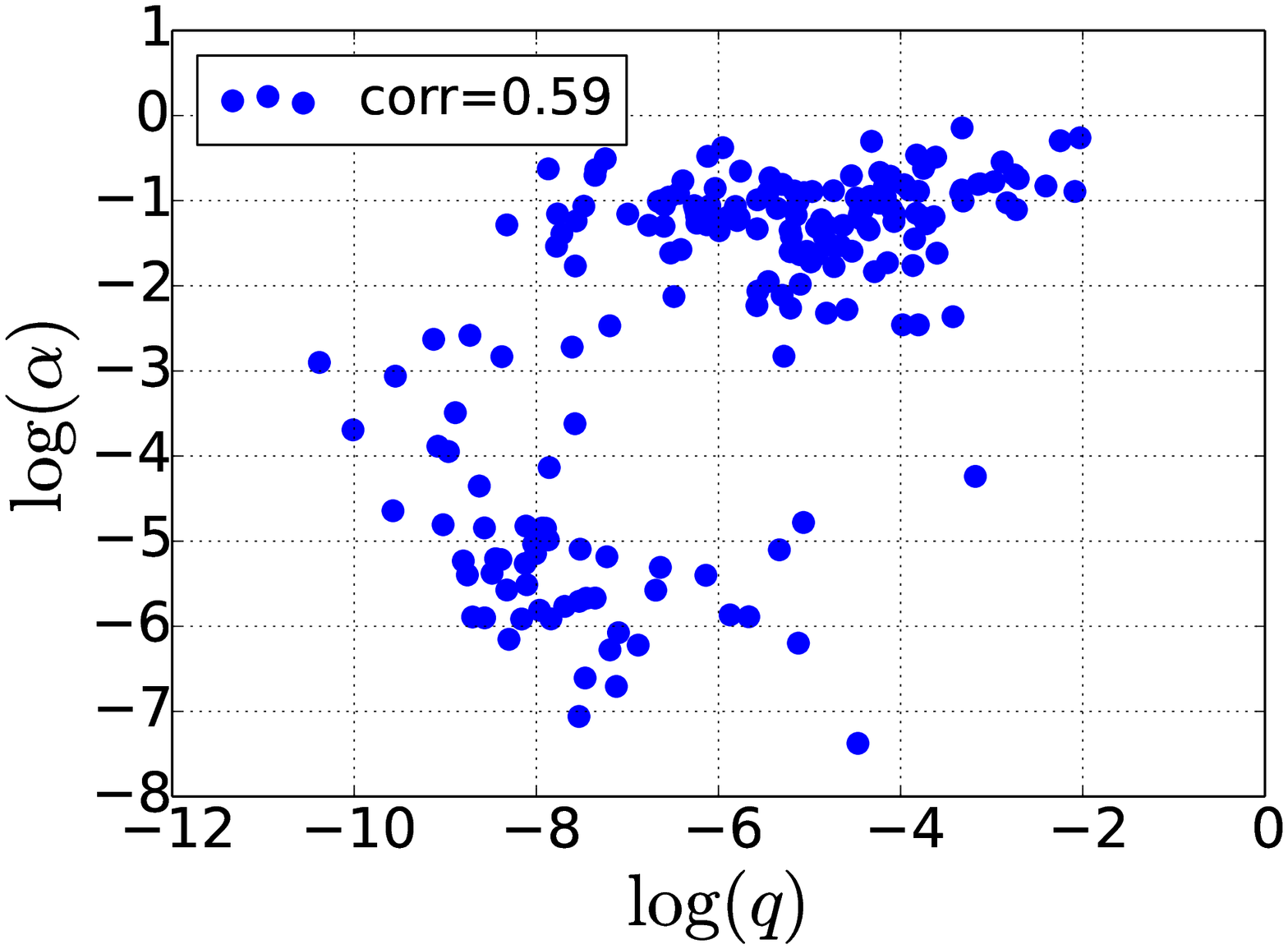}
                \caption{}
                \label{fig:cstv2}
        \end{subfigure}
        \begin{subfigure}[b]{0.30\linewidth}
                \includegraphics[width=1.0\linewidth]{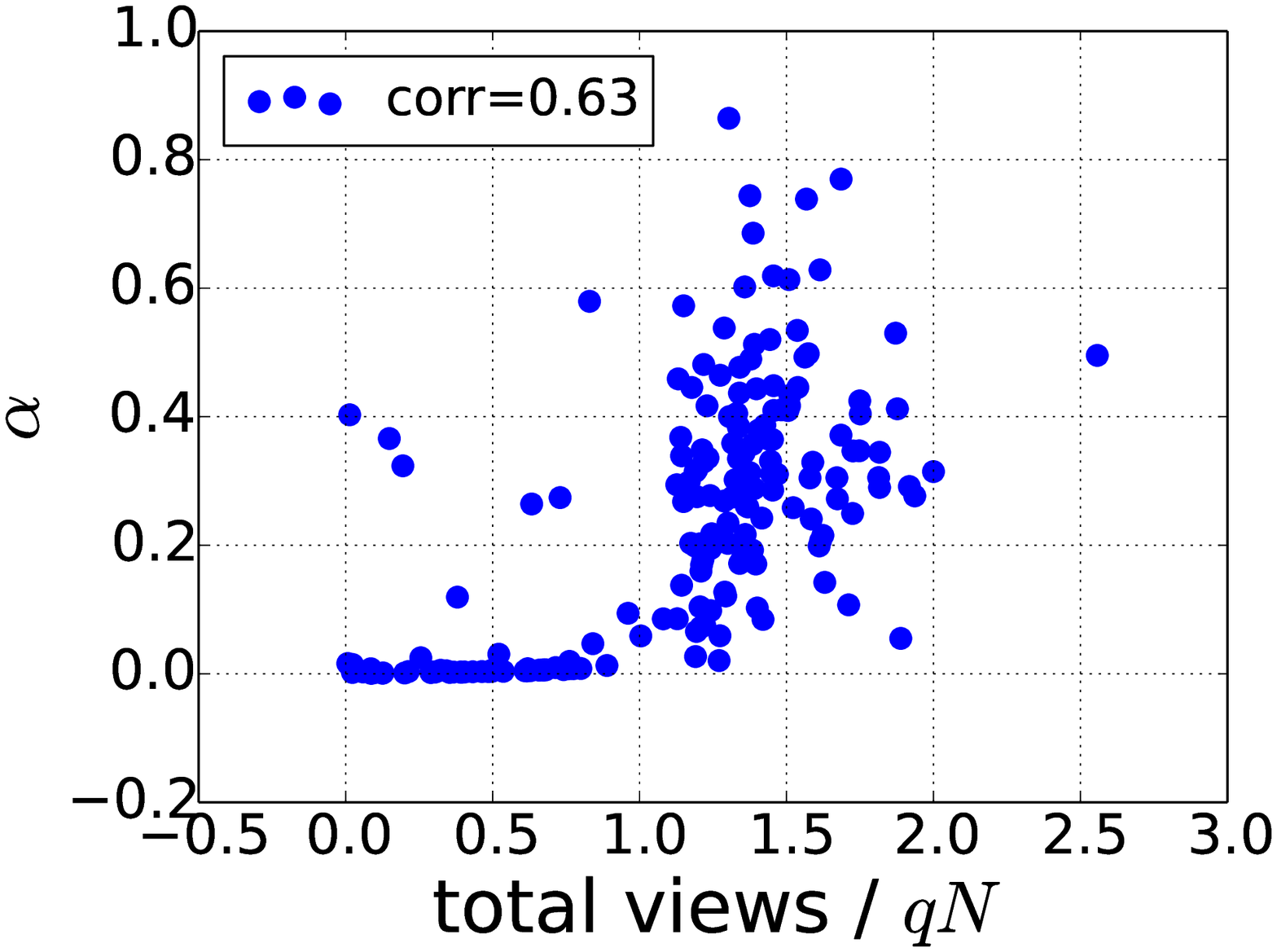}
                \caption{}
                \label{fig:cstv3}
        \end{subfigure}

                \begin{subfigure}[b]{0.30\linewidth}
                \includegraphics[width=1.0\linewidth]{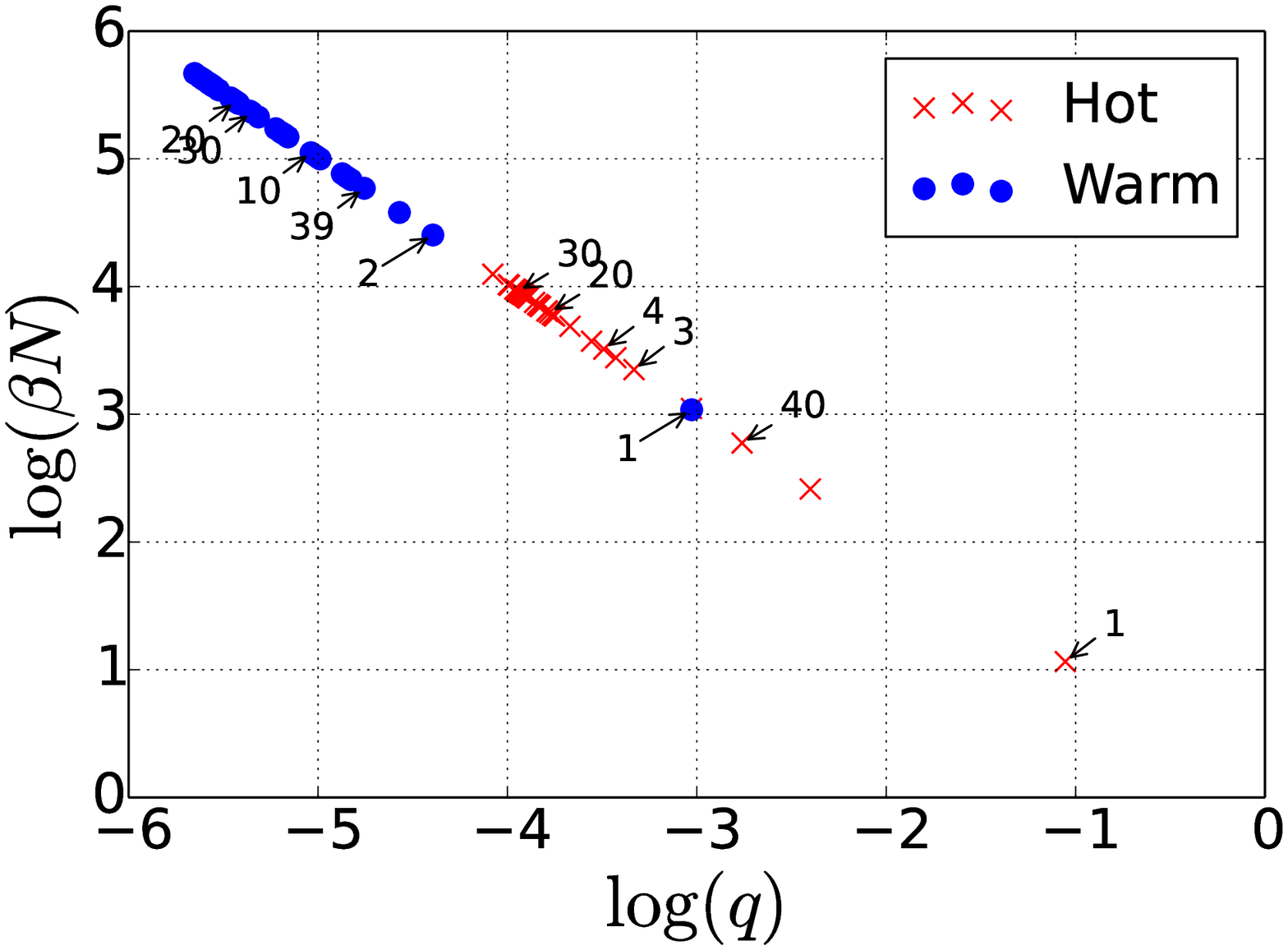}
                \caption{}
                \label{fig:cstv4}
        \end{subfigure}
        \begin{subfigure}[b]{0.30\linewidth}
                \includegraphics[width=1.0\linewidth]{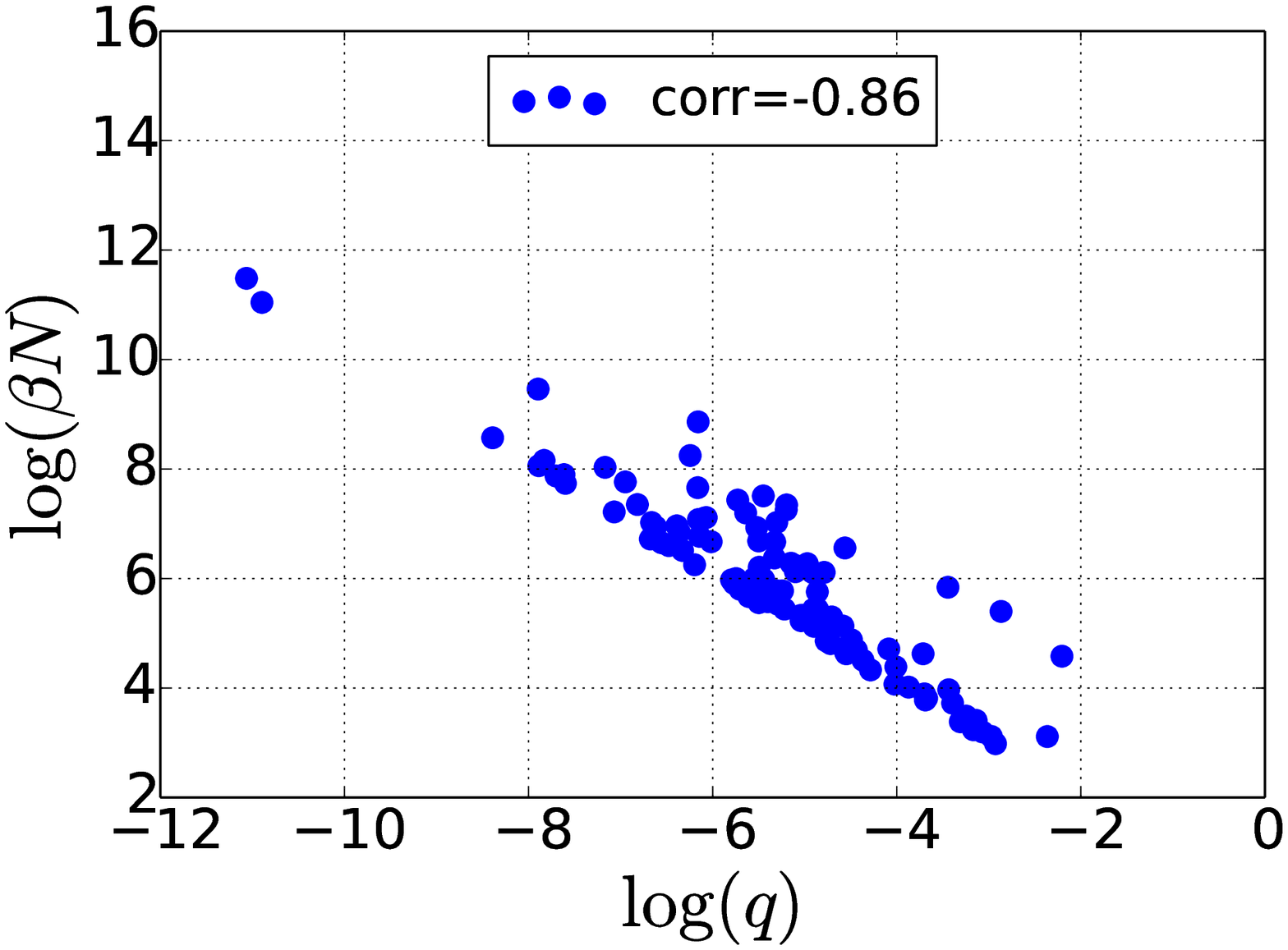}
                \caption{}
                \label{fig:cstv5}
        \end{subfigure}
        \begin{subfigure}[b]{0.30\linewidth}
                \includegraphics[width=1.0\linewidth]{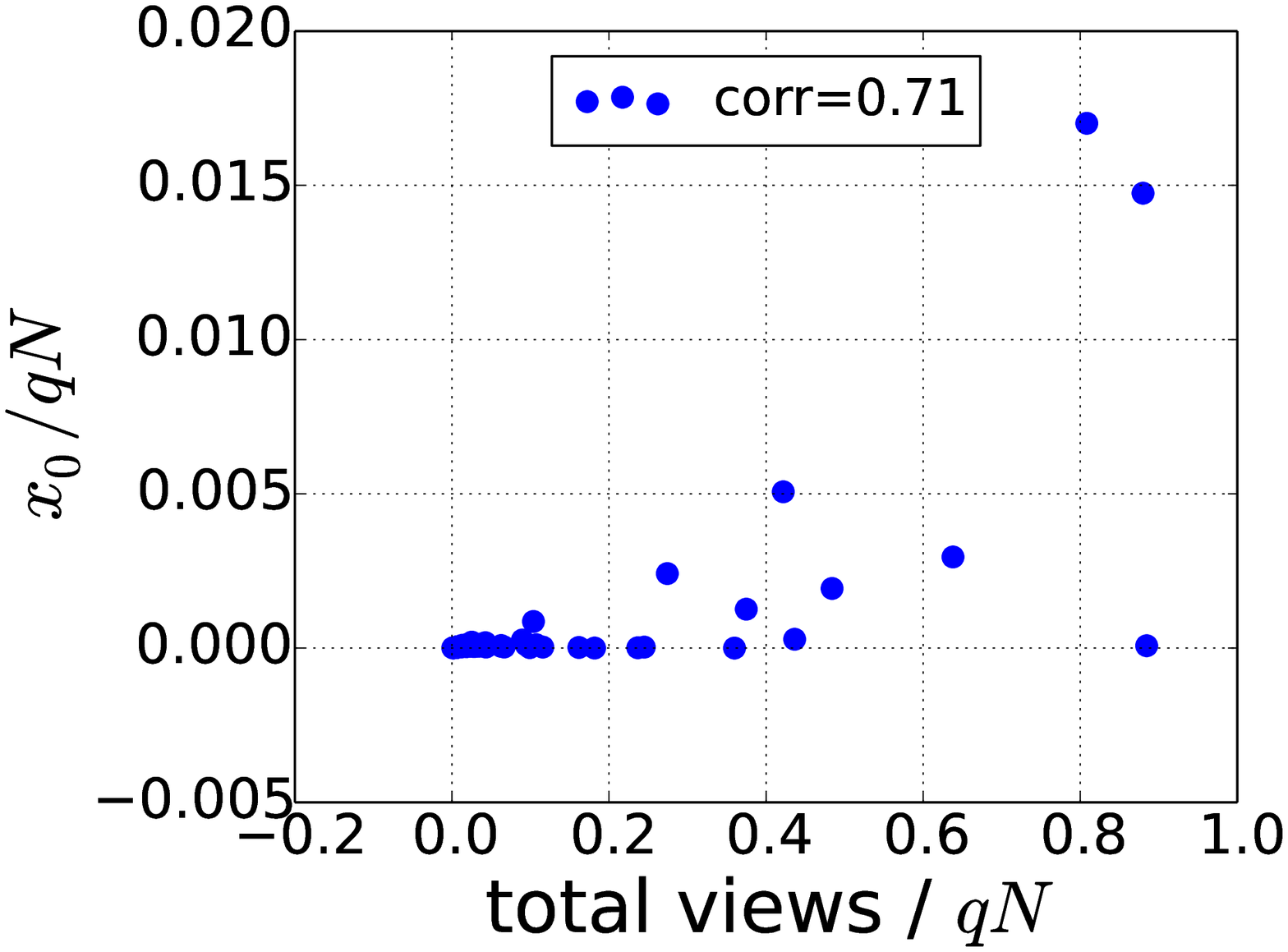}
                \caption{}
                \label{fig:cstv6}
        \end{subfigure}

        \caption{User behavior patterns and recommendation strategies.}
        \label{Fig:cstv}
\end{figure*}
Both DModel and WModel assume that users make independent decision to view videos. In reality, some videos may be related such as the News videos about the same event and different episodes of the same TV series. Thus, it is reasonable and necessary to aggregate these related videos into a single \emph{composite} video before we study the recommendation strategies.

There are a total number of 9010 TV episodes used for case study, after excluding 695 badly fit TV episodes with large NMSE. For each TV series, we aggregate all episodes and different versions (e.g, high/low definition versions, different language versions) into a composite video by averaging their parameters $\alpha$, $\beta$, etc. Each composite video is classified as a d-recommended video if most of its episodes achieve better fitting with DModel, or a W-recommended video otherwise. After aggregation, there are $190$ d-recommended and $112$ w-recommended composite videos. Most d-recommended videos are newly produced TV series while most w-recommended videos are old TV series which are uploaded to the system recently. They are investigated separately below because of their different recommendation strategies and user behavior patterns.

\subsection{D-recommended videos}
Fig.~\ref{fig:cstv1} shows the scatter plot of $\log(q)$ v.s. $\log(\alpha)$ of all the episodes from two sampled TV series \emph{before aggregation}. The cross dots represent the parameters of a hot TV while the round dots represent a warm TV. Some selected dots are annotated with episode numbers. We can observe that the first episodes attract more views than the following episodes for both TV series. As episode number increases, attractiveness decreases gradually, indicating that the first episode is the most attractive one and not all users can keep up with the update of TV episodes.

Fig.~\ref{fig:cstv2} shows the scatter plot of $\log(q)$ v.s. $\log(\alpha)$, and Fig.~\ref{fig:cstv3} shows the scatter plot of $\frac{\text{ view count}}{ qN}$ v.s. $\alpha$ for the 190 composite videos. Both scatter plots show strong positive Pearson correlations. Note that users are likely to make independent decisions to view composite videos, and thus it is fair to compare recommendation resources allocated to composite videos. Fig.~\ref{fig:cstv2} indicates that a strong positive correlation exists between video intrinsic attractiveness and direct recommendation strength. Concentrating on recommending popular videos is a reasonable strategy given limited direct recommendation resource. Fig.~\ref{fig:cstv3} shows large $\alpha$ tends to achieve high completion fraction, confirming that video information will be diffused more sufficiently if it is allocated with more recommendation resource.

\subsection{W-recommended videos}
Similar to Fig.~\ref{fig:cstv1}, we investigate the relationship between intrinsic popularity $q$ and its WOM recommendation rate $\beta$ for episodes within two sampled TV series. Fig.~\ref{fig:cstv4} is the scatter plot of $\log(q)$ and $\log(\beta N)$ of all the episodes from a hot TV series with 40 episodes and a warm TV series with 39 episodes. An interesting observation is that both the first several episodes and the last several episodes are very popular. Such different user behaviors from Fig.~\ref{fig:cstv1} are due to the fact that episodes of d-recommended TVs are uploaded slowly, typically one or two episodes per day. However, all episodes of most w-recommended TV series are uploaded together because they are produced several months or years ago. Given all episodes available, users are prone to browse the first and the last several episodes.

For episodes in Fig.~\ref{fig:cstv1} and  Fig.~\ref{fig:cstv4}, there is a trend that recommendation rate increases with episode number. We believe that earlier episodes act as ``advertisements" of the later episodes, such that user reaction rate becomes faster and faster. An exception is the last several episodes of w-recommended videos. We believe that these videos are browsed just following the browsing of the first several videos.

Fig.~\ref{fig:cstv5} shows the scatter plot of $\log(q)$ v.s. $\log(\beta N)$ and Fig.~\ref{fig:cstv6} shows the scatter plot of $\frac{\text{ view count}}{ qN}$ v.s. $\frac{x_0}{qN}$ for the 112 w-recommended composite videos. Surprisingly, a strong negative correlation appears between $\log{\beta N}$ and $\log{q}$ in Fig.~\ref{fig:cstv5}, indicating that users prefer recommending less popular videos. This observation is consistent with our intuition. People like to share interesting but uncommon things with friends. Fig.~\ref{fig:cstv6} shows a strong positive correlation between $\frac{\text{view count}}{qN}$ v.s. $\frac{x_0}{qN}$, implying that initial seed population is very essential for information diffusion.

%% file: sec/related.tex
\section{Related Work}
Some previous studies have proposed several models of dynamic video popularity.
In \cite{liuNOSSDAV12}, the authors studied video sharing in online social network and proposed a probabilistic model that matches the observed dynamic popularity distribution. But their model does not depict individual videos' popularity evolutions. Avramova et al.~\cite{avramovaICEI09} proposed a closed-form expression for a video's popularity evolution, which can be degenerated into either a power-law or exponential decay function. However, it is an ad-hoc model and does not reveal information diffusion process.
In \cite{jasonIWQOS15}, although the authors consider both recommendation mechanisms, their model is rather complicated and cannot be easily applied to quantify the recommendation strength.

There exist works that studied popularity prediction. Szabo and Huberman~\cite{szaboCACM10} found that a linear relationship exists between a video's early and future popularity, and thus future popularity can be predicted through multiplying early popularity by a linear coefficient learned from past datasets. In \cite{pintoWSDM13}, the authors generalized the Szabo-Huberman method by considering multiple early popularity and video similarity. Ahmed et al.\cite{ahmedWSDM13} classified videos by their popularity evolution patterns and predicted future popularity based on the classification. These predictive models focus on prediction accuracy but lack explainable and measurable factors, hence cannot give theoretical insights.

Video popularity is an important reference for resource allocation in content delivery. Researchers have studied various schemes by incorporating popularity dynamics. For example, Zhou et al. \cite{yipengTMM15} proposed a mixed CDN caching strategy for both age-sensitive and popularity-stable videos. Hu et al \cite{huICME2014} leveraged community based social video viewing behaviors to develop a cloud CDN content placement scheme. \cite{alexICME2013} tried to detect popular videos as CDN caching candidates by mining main steam media. However, in these works the popularity analysis are mainly based on measurements and theoretical models are not proposed.

%% file: sec/conclusion.tex
\section{Conclusion}
In this paper, we develop a dynamic model to depict video popularity evolution. Specifically, two recommendation mechanisms, direct and WOM recommendations, are incorporated in the model as the driving forces of video information diffusion process. Our model provides a systematic approach to quantifying video recommendation forces and other factors, and is useful for characterizing user behaviors and evaluating recommendation strategies, which are illustrated with a case study of TV episodes. Extending our model to involve more factors and predicting video popularity evolution will be our future work.

%% file: sec/appendix.tex
\section*{Appendix}
\noindent{\bf Derivation of Eq.~\ref{EQ:xt_beta}}

Let $x^{'}=p$ and thus we have $x^{''} = p^{'}$. Since
$$
p=\frac{dx}{dt}=\frac{dx}{dp}\cdot \frac{dp}{dt} = \frac{dx}{dp}\cdot p^{'}
$$
, we have $p^{'}=p\cdot \frac{dp}{dx}$.

Note that $s=N-\frac{x}{q}$. Substituting it back to Eq.~\ref{EQ:dx_beta_cont}, we have
\begin{eqnarray*}
  &&  x^{''} = x^{'} (\beta qN - \beta x - 1) \\
   &\Rightarrow& p \cdot \frac{dp}{dx} = p(\beta qN - \beta x - 1) \\
   &\Rightarrow& \frac{dp}{dx} = \beta qN - \beta x - 1 \\
   &\Rightarrow& p = (\beta qN - 1) x - \frac{1}{2}\beta x^2 + C \\
\end{eqnarray*}
,where $C$ is a to-be-determined constant. With the initial condition $x^{'}(0) = \beta qx_0(N-\frac{x_0}{q})$, we can solve for $C$ and
the two roots of $- \frac{1}{2}\beta x^2 + (\beta qN-1) x + C=0$ are
\begin{eqnarray*}
  x_1 &=& qN+\frac{\varphi-1}{\beta} \\
  x_2 &=& qN+\frac{-\varphi-1}{\beta}
\end{eqnarray*}
,where $\varphi = \sqrt{(\beta qN-1)^2-(\beta x_0-1)^2+1}$. Thus we have
\begin{eqnarray*}
  &&  \frac{dx}{dt} = - \frac{1}{2}\beta x^2 + (\beta qN-1) x + C \\
   &\Rightarrow& \int \frac{dx}{- \frac{1}{2}\beta x^2 + (\beta qN-1) x + C} = \int dt \\
   &\Rightarrow& \int \frac{1}{-\frac{1}{2}\beta(x_1 - x_2)} \cdot \big(\frac{1}{x-x_1} - \frac{1}{x-x_2}\big) dx = \int dt \\
   &\Rightarrow& \frac{1}{-\varphi} \cdot \ln{|\frac{x-x_1}{x-x_2}|} = t + D
\end{eqnarray*}
It is not difficult to show that $x_2 < x_0 < x_1$. Therefore,
$$
\frac{x-x_1}{x - x_2} = -g(t) \Rightarrow x(t) = \frac{x_1+g(t)x_2}{1+g(t)}
$$
. With the initial condition $x_0$, we can get $g(t) = \frac{x_1-x_0}{x_0-x_2}e^{-\varphi t}$.

\done
\noindent{\bf Proof of Proposition~\ref{Prop:x1_beta}:}
Let $f(x_0, \beta) = x_1 = qN + \frac{\varphi - 1}{\beta}$, where $\varphi = \sqrt{(\beta qN-1)^2-(\beta x_0 - 1)^2+1}$. Since
$$
\frac{\partial \varphi}{\partial \beta} = \frac{(qN-x_0)(\beta qN + \beta x_0 - 1)}{\varphi}
$$, we have
\begin{eqnarray*}
  \frac{\partial f}{\partial \beta} &=& \frac{\beta(qN-x_0)+\varphi-1}{\varphi \beta^2} \\
   &=& \frac{\beta(qN-x_0)+\beta (x_1 - qN)}{\varphi \beta^2} \\
   &=& \frac{x_1-x_0}{\varphi \beta} > 0
\end{eqnarray*}
Hence, final population $x_1$ always increases as $\beta$ increases.
\begin{eqnarray*}
  \frac{\partial^2 x_1}{\partial \beta^2} &=& \frac{\frac{\partial x_1}{\partial \beta}\varphi \beta - (x_1-x_0)(\frac{\partial \varphi}{\beta}\beta+\varphi)}{\varphi^2\beta^2},\\
  & = & \frac{(x_1-x_0)}{\varphi^2\beta^2}\left(1-\frac{\partial \varphi}{\beta}\beta-\varphi\right)
\end{eqnarray*}
$x_>x_0$. By letting $\left(1-\frac{\partial \varphi}{\beta}\beta-\varphi\right) = 0$, we derive $\beta = \frac{1}{2(qN+x_0)}$.

\done